# Dynamics of a superconducting linear slider

Ignacio Valiente-Blanco*[1], Jose-Luis Perez-Diaz[2] and Efren Diez-Jimenez[2]

[1]*Instituto Pedro Juan de Lastanosa, Avenida de la Universidad 30, E-28911 Leganés, Spain*

[2] *Dto. de Ingeniería Mecánica, Universidad Carlos III de Madrid, Butarque, 15, E-28911 Leganés, Spain*

**Abstract**

In this paper, the dynamic behavior of a one DoF contactless linear slider based on superconducting magnetic levitation is experimentally analyzed. The device is intended for precision positioning of an optic mirror in cryogenic environments. Different prototypes of this device have been tested at cryogenic temperatures (77 K), and their mechanical behavior characterized in the sliding direction for forced and unforced oscillations. Experimental results reveal that the slider is self-stable at the initial equilibrium position and the dynamic behavior fits well an underdamped harmonic oscillator. Finally, the device showed great potential for horizontal vibration isolation, acting as a low-pass filter with a resonance at about 0.9 Hz.

* Corresponding author: Tel.: +34 916249912; fax: +34 916249912.

E-mail address: ivalient@ing.uc3m.es

1. **Introduction**

The applications of classical mechanical joint and mechanisms to cryogenic applications present several limitations, such as wear, cold welding, the limited number of options for lubrication at cryogenic temperatures and the presence of hysteresis, all of which are mainly induced by contact between moving parts [1–3]. These limitations can be overcome by contactless mechanisms such as active magnetic levitation systems, which are frequently used as dampers, vibration isolators and precise positioning devices among other applications [4–8]. In addition, superconducting magnetic levitation has been demonstrated to be an alternative for the design of contactless mechanisms. Its intrinsic stability [9,10] is an important advantage over active magnetic levitation, which is inherently unstable as introduced by Earnshaw's theorem [11] and typically require more complex control strategies, higher power consumptions and have reduced reliability [12,13]. Due to this list of advantages, superconducting devices has made possible to build flywheels for energy storage [14–16], superconducting magnetic bearings [17,18], vibration isolators for space structures [19] and precise positioning devices [20].

Precision positioning, both at room temperature and in cryogenic environments, has become an important development for precision mechanisms and device: refocusing mechanisms, optical path difference actuators (OPDA), cryogenic motors, multi-axis positioning stages, conveyors and actuators [1–4][5], [6] are some examples of application. Infrared interferometer spectroscopy [7] is an application of particular interest for this device [8]. In previous works we presented a new concept of a superconducting non-contact linear slider for precision positioning of a optic mirror cube as part of the mechanisms required in far infrared interferometry [21,22]. In addition, the slider can also be used as a low-maintenance, high-reliability sliding kinematic joint for space application due to the lack of contact and its self-stability [23,24]. In this context, the dynamic behavior is a major issue to be considered for the design and control of this sort of mechanism. In spite of the stability of superconducting mechanisms, their dynamics have been found to be relatively complex [25–30], even chaotic [31]. In the present paper we experimentally show that the dynamics of that device fits reasonably well to an underdamped harmonic oscillator with a natural frequency about 1 Hz.

**2. Description of the device**

The device studied in this paper is a slider mechanism for precision positioning in cryogenic environment. It is mainly composed of a static guideline, made of two 45 mm diameter superconducting polycrystalline $YBa_2Cu_3O_{7-x}$ disks (HTS) (1), and a slider composed of a long $Nd_2Fe_{14}B$ permanent magnet (PM) (2). The permanent magnet is a bar 160 mm in length and with a square section of 10x10 mm$^2$ with a remanence $Br$ = 1.3T and a coercivity $Hc$ = 900 kA/m. Note that the magnetization direction is parallel to the Z axis in Fig. 1. The remaining components in the device are made of aluminum, with good thermal conductivity and low magnetic interaction. In previous works we have demonstrated that once the superconductors are at a temperature below their critical temperature (T= 93K) [32], the slider levitates stably over the HTS and presents an equilibrium position at the center of the reference system (Fig. 1) [23,33]. In addition, it has also been demonstrated that the slider present a much higher resistance to the motion in any other direction but the sliding direction due to the magnetic fields applied from the PM to the HTS and it is self-stable in all directions.

**INSERT FIG.1**

It has been demonstrated that the performance of the mechanism can be easily tuned by geometric modifications of the prototype [24]. In this sense two parameters have a major role. One of them is the distance between the superconductors in the stator, d, and the other is the height of levitation of the PM or height of field cooling (HFC), defined as the distance between the superconductors's top surface and the PM bottom surface, both represented in Fig. 2. The influence of the value of *d* in the selection of the natural frequency of the device will be discussed in the following sections.

The central equilibrium position can be easily relocated along the stroke of the mechanism by modifying, with an open-loop control strategy, the current in two coils placed at either end of the stroke of the mechanism (3). The axes of the coils are parallel to the magnetization axis of the PM. Finally, these coils have been specially designed to provide improved stability to the PM and only exert distance magnetic forces in the sliding direction.

**3 Experimental study**

*3.1 Experimental set up and procedure*

The experiments have been developed using the experimental set-up presented in Fig. 2. The prototype has been contained in a vessel (8) and immersed in a liquid nitrogen bath at ambient pressure (~77 K). Then, the temperature of the superconductors is reduced to 77.3 K, well below the critical temperature at which they become superconducting ($T_c$= 93K). Both superconducting disks (1) have been located at different distances, d, using different housings. The value of d has been modified between 47 and 90 mm. In addition, the height of field cooling (HFC) or levitation height has been fixed at 3± 0.1 mm in all cases, using a pair of spacers. For damping, speed and transmissibility analysis in sections 3.3 to 3.5 respectively, a prototype with d= 84 mm and HFC= 3 mm has been selected.

The actuation system is composed of a couple of coils (3) with rectangular shape (110x60 mm2) and 10 mm in height with 190 AWG18 wire turns each located at each end of the stroke and connected in anti-serial configuration unless other is specified. This connection present the advantage that while one of the coils repels the PM, the other attracts it therefore increasing the magnitude of the force exerted and reducing the power consumption with regard to other configurations. The measured resistance of each coil at room temperature is about 3.8 Ω.

Once a configuration of the prototype is selected (d and HFC) the superconductors are cooled and, after a short period of time for the stabilization of the temperature, the spacers removed leaving the PM able to levitate. After levitation has been established, the coils (3) were supplied with a DC or AC current provided either by a Velleman DC power supply model ps613 or a Promax function generator model gf-855. The DC power source has been used to provide the current in the coils for the motion of the slider in sections 3.2 to 3.4. The AC signal generator has been used in order to study the dynamic response of the slider under sinusoidal forced oscillations, introduced in section 3.5. In all cases, the current in the coils has been measured using a multimeter from ICE Instrumentazione, model 5600, with 0.01 mA accuracy.

In order to measure the position of the slider in its path, a laser triangulator ILD 1402 from Micro-Epsilon, with a resolution of about 10 μm, has been used (4). The slider carries an aluminum cube (5) on which the laser beam is reflected. The height of the prototype with respect to the beam from the laser triangulator can be regulated thanks to a lab-jack stand (6). Finally, the whole set-up is installed onto an optic table (7).

**INSERT FIG.2**

*3.2 DC Current in the coils vs. position of the slider*

The X position of the slider vs. the DC current in the coil is plotted for different values of *d* in Fig. 3. For this experiment only one coil was acting, and always with a polarity so that it repels the PM. Induced displacement in the Y and Z axes was below the measurement system resolution (250 μm) for any position of the slider in the stroke.

**INSERT FIG.3**

X=0 mm is identified as a stable equilibrium point of the system. It is clear from Fig. 3 that the higher the value of d, the higher the stiffness and the lower the sensitivity to the motion of the slider in its path. This is assumed to be caused by the increased component of the magnetic field in the sliding direction seen by the superconductors for higher values of *d*, which increases the force exerted by the superconductors on the PM [23], [24].

*3.3 Damping and natural frequency*

In order to characterize the damping and natural frequency of the mechanism for unforced oscillations, the following experiment has been developed: initially, the coils are supplied with a constant DC current and the slider moves to the limit of the stroke (X ~ +10mm). Then, once the position has been stabilized, the coils were switched off and the position of the slider was measured until it stabilizes its position at the equilibrium point of the system (X = 0). The results of this experiment are presented in Fig.4.

**INSERT FIG.4**

The damping of the system can obtained from Fig. 4 using the logarithmic decrement method and it is about $\xi = 0.18 \pm 0.01$. We assume that the damping of the slider is mainly caused by eddy current in the structural parts of the parts of the device and also energy dissipation in the superconductors. It is well known that the presence of impurities and defects in the superconductor reduces the mobility of the magnetic vortexes in it [33]. Air drag probably has its influence too, but the contribution of each of them has not been characterized separately.

The dashed line in Fig. 4 represents the response of a harmonic oscillator with $M = 185\ g$; $\omega_d = 5.6\frac{rad}{s}$ and $\xi = 0.18$. From Fig. 4, it is evident that the mechanism behaves as an underdamped oscillator. The damped frequency of the system can be determined by time analysis from the data plotted in Fig. 4 or using the Lomb-normalized periodigram method. This technique can be extended to situations in which the samples are taken at uneven intervals (as in this case). In addition, this method will reveal the presence of harmonics in the oscillation, if present. The results of the Lomb-normalized periodigram analysis are plotted in Fig. 5.

**INSERT FIG.5**

No significant harmonics are appreciated in Fig. 5 for frequencies below 100 Hz. The damped frequency has been obtained at f = 0.9±0.1 Hz (time analysis of data in Fig.4 gives a damped frequency of 0.91±0.03 Hz). From the damped frequency, the natural frequency of the system can then be calculated as:

$$f_0 = \frac{f_d}{\sqrt{1-\xi^2}}$$

eq. 1

Natural frequency from experimental data is about 0.92±0.04 Hz in very good agreement with which it was predicted in section 3.2, supporting the idea that the behavior of the slider can be described with good accuracy by an underdamped harmonic oscillator.

*3.4 Speed of the slider*

The speed of the slider has been derived from the position vs. time data in Fig. 4. Results are plotted in Fig.6 . The equilibrium position of the mechanism can be seen in the attractor in Fig. 6. Ultimately, speeds of up to 30 mm/s were measured for an unforced oscillation of amplitude 10 mm (near the limit of the stroke). Black dashed line in Fig. 6 represents the ideal trajectory of a harmonic oscillator.

**INSERT FIG.6**

*3.5 Transmissivity analysis*

The motion of the slider for forced harmonic oscillations has been studied with the following procedure. A sinusoidal current of constant magnitude (about 90 mA) was supplied to both of the coils in the mechanism. These coils were connected in anti-serial configuration, so that when one of the coils attracts the PM the opposite one repels it. This connection is intended to increase the sensitivity of the motion of the slider with respect to the configuration adopted in section 3.2 and simplify the control of the position of the mechanism [24]. Despite the magnitude of the current being constant (and also being the amplitude of the excitation force on the slider), the frequency of the excitation was modified between 0.01 Hz and 100 Hz. Then, the amplitude of the resulting oscillation is compared with the reference value of the displacement for a continuous current signal of 90 mA amplitude (reference displacement about 2.3 mm).

Transmissibility can be calculated from experimental data and compared to the transmissibility expected in the corresponding harmonic oscillator using eq. 2.

$$T = \left|\frac{A_o}{A_i}\right| = \sqrt{\frac{1 + (2\xi\Omega)^2}{(1 - \Omega^2)^2 + (2\xi\Omega)^2}}$$

eq. 2

where

$A_o$ is the amplitude of the response,

$A_i$ is the amplitude of the vibrational input: Displacement amplitude for f ~0 Hz equal to 2.3 mm,

$\Omega = \frac{\omega}{\omega_o}$ is the frequency ratio,

and $\xi$ is the damping ratio.

Results of these measurements are plotted in Fig. 7.

**INSERT FIG.7**

No further results for frequencies higher than 100 Hz are presented, because the amplitude of the transmitted oscillation is less than the displacement sensor resolution. However it seems clear that the device behaves like a low-pass filter for vibrations in the horizontal direction, with its resonant frequency at about 0.9 Hz. Besides, the transmissibility of the device can be described with reasonable accuracy by a harmonic oscillator.

## 4. Conclusions

The dynamic behavior of a levitating superconducting linear slider is presented. A set of experiments at 77 K and ambient pressure has been developed with different prototypes. Despite dynamic behavior of plenty of superconducting devices in literature has been found complex, even chaotic, experimental analysis on the device presented in this paper show that the motion of the slider in the horizontal direction can be described with good accuracy by an underdamped harmonic oscillator with damping linearly dependent on the speed of the slider. Results for transmissibility experimental analysis for sinusoidal forced oscillations also support the idea that the slider behaves like a harmonic oscillator. Ultimately the characteristic parameters of the oscillator that have been experimentally characterized are summarized in Table 1.

**INSERT Table I.**

In addition, the device showed the potential to become a contactless passive or active system for horizontal vibration isolation with a low natural frequency below 1 Hz with a quality factor about 3. Finally, results obtained in this study will allow engineers to design different devices for vibration isolation and precision positioning in cryogenic environments.

**Acknowledgement**

This work has been partially funded by the Dirección General de Economía, Estadística e Innovación Tecnológica, Consejería de Economía y Hacienda, Comunidad de Madrid, ref. 12/09. The authors would like to thank Javier Serrano, F. Romera, H. Arguelaguet-Vilaseca and D. González-de-María from LIDAX for their technical support and assistance. In addition, we would like to thank Juan Sánchez-García-Casarrubios his cooperation in the project.

**Figure caption listing**

| | |
|---|---|
| **Fig. 1.** | Picture of the device: (1) YBaCuO superconductor disks; (2) Slider permanent magnet; (3) Coils, (4) Optic mirror cube. |
| **Fig. 2**. | Sketch of the experimental set-up: (1) YBaCuO superconductor disks; (2) Slider, permanent magnet; (3) Coils; (4) Laser triangulator ILD 1402; (5) Polished aluminum mirror cube; (6) Lab-jack stand; (7) Optic table; and (8) Liquid nitrogen vessel. *d*: distance between the superconducting disks and *HFC* = height of field cooling. |
| **Fig. 3**. | Position X vs. DC current in the coil for different values of *d*. T= 77 K and *HFC*= 3 mm in all cases. |
| **Fig. 4**. | Position X vs. time for an unforced oscillation of the slider. T = 77 K, d = 84 mm and HFC = 3 mm. Reference amplitude of the oscillation about 10 mm. |
| **Fig. 5**. | Power spectrum vs. frequency of the Lomb-normalized periodigram of the signal in Fig. 4. |
| **Fig. 6**. | Speed vs. Position X of the slider is represented by grey line. The ideal response of a harmonic oscillator with $\xi$=0.18 and $\omega_0$=0.93 is represented by the black dashed line. |
| **Fig. 7**. | Transmissibility vs. frequency ratio. Displacement amplitude for f ~0 Hz is approximately 2.3 mm. Dashed line represent transmissibility for a harmonic oscillator with $\xi$=0.18 and $\omega_0$=0.93 Hz. |

**Table caption listing**

   **Table 1**.        Summary of the dynamic parameters of the slider

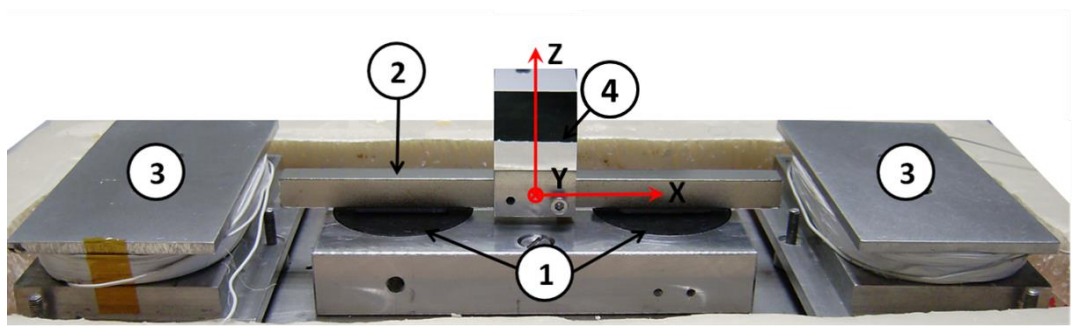

**Fig. 1.** Picture of the device: (1) YBaCuO superconductor disks; (2) Slider permanent magnet; (3) Coils, (4) Optic mirror cube.

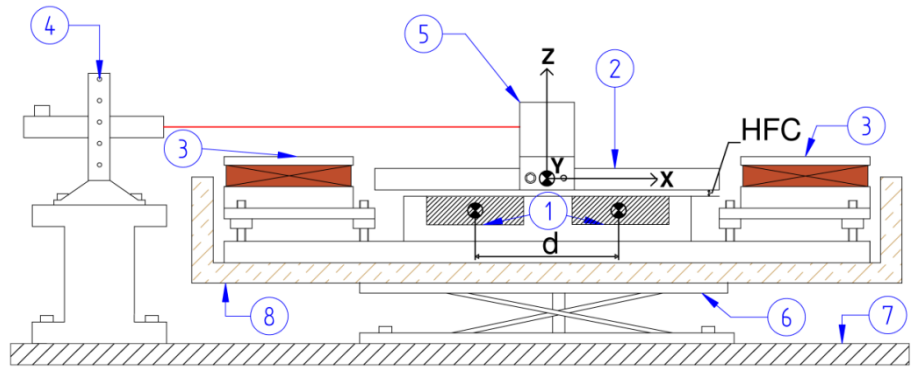

**Fig. 2**. Sketch of the experimental set-up: (1) YBaCuO superconductor disks; (2) Slider, permanent magnet; (3) Coils; (4) Laser triangulator ILD 1402; (5) Polished aluminum mirror cube; (6) Lab-jack stand; (7) Optic table; and (8) Liquid nitrogen vessel. *d*: distance between the superconducting disks and *HFC* = height of field cooling.

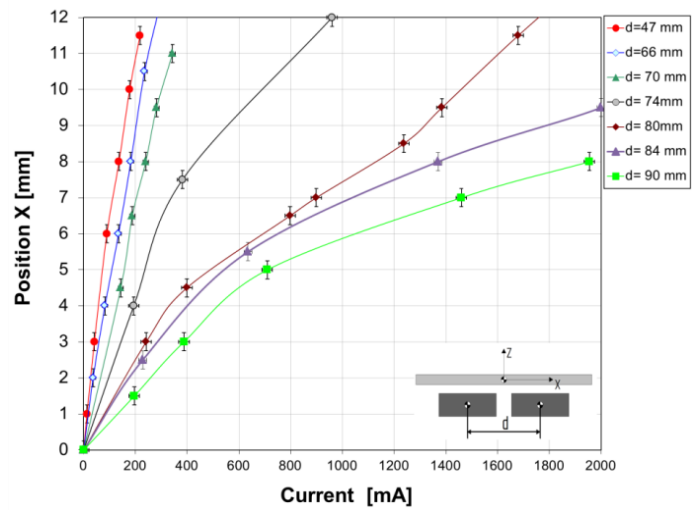

**Fig. 3**. Position X vs. DC current in the coil for different values of *d*. T= 77 K and *HFC*= 3 mm in all cases.

*Journal of Vibration and Acoustics*

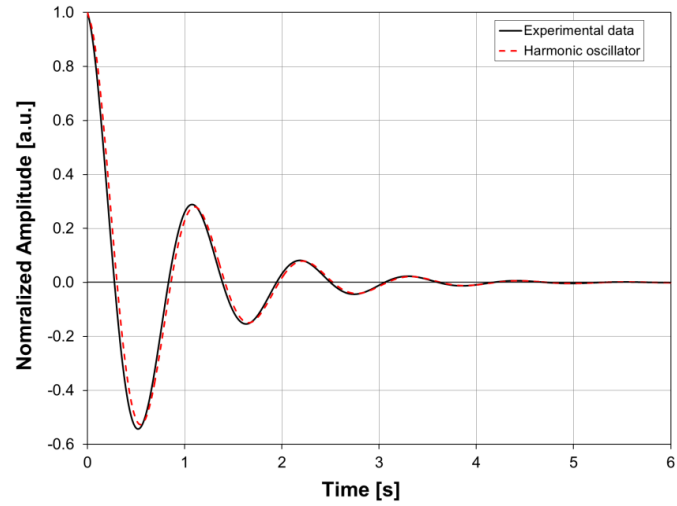

**Fig. 4**. Position X vs. time for an unforced oscillation of the slider. T = 77 K, d = 84 mm and HFC = 3 mm. Reference amplitude of the oscillation about 10 mm.

*Journal of Vibration and Acoustics*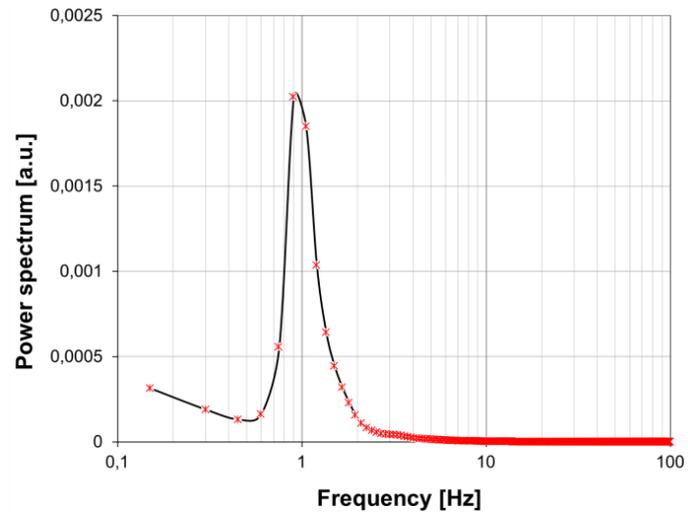

**Fig. 5**. Power spectrum vs. frequency of the Lomb-normalized periodigram of the signal in Fig. 4.

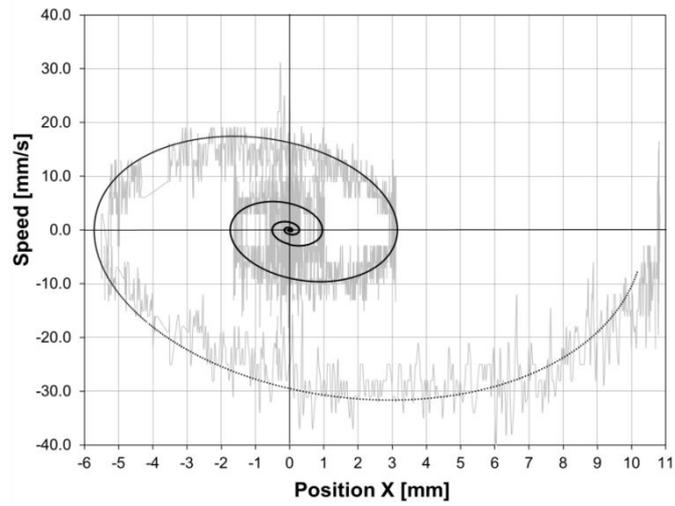

**Fig. 6**. Speed vs. Position X of the slider is represented by grey line. The ideal response of a harmonic oscillator with ξ=0.18 and ω$_0$=0.93 is represented by the black dashed line.

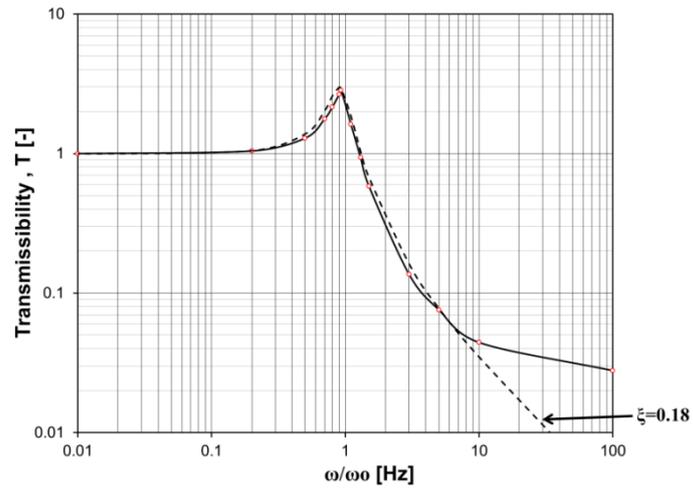

**Fig. 7**. Transmissibility vs. frequency ratio. Displacement amplitude for f ~0 Hz is approximately 2.3 mm. Dashed line represent transmissibility for a harmonic oscillator with ξ=0.18 and $\omega_0$=0.93 Hz.

**Table 1**. Summary of the dynamic parameters of the slider

| Parameter | Value |
|---|---|
| Experimental temperature [K] | 77 |
| Mass of the slider [g] | 185±5 |
| Natural frequency [Hz] (experimental) | 0.92±0.04 |
| Natural frequency [Hz] (FEM) | 0.91±0.04 |
| Damping factor, ξ [-] | 0.18±0.01 |
| $c_c$ [Nsm$^{-1}$] | 2.1±0.2 |
| $c$ [Nsm$^{-1}$] | 0.38±0.06 |
| Maximum velocity (unforced oscillation) [mm·s$^{-1}$] | ≈ 30 mm/s |